\documentstyle[11pt,newpasp, twoside, epsf]{article}   
\def\etal{{\it et al.~}}

\def\edcomment#1{\iffalse\marginpar{\raggedright\sl#1\/}\else\relax\fi}
\reversemarginpar

\begin{document}
\title{Fragmentation and Star Formation in Turbulent
Cores}
\author{Richard I. Klein}
\affil{University of California,
Lawrence Livermore National Laboratory and Berkeley Department of
Astronomy, 601 Campbell Hall, Berkeley, California, 95720}
\author{Robert Fisher}
\affil{University of California, Berkeley, Department of Physics}
\author{Christopher F. McKee}
\affil{University of California, Berkeley, Department of Physics}

\begin{abstract}

We  examine the conditions under which binary and multiple stars may
form out of turbulent molecular cloud cores using
 high resolution 3-D, adaptive mesh refinement (AMR) hydrodynamics
(Truelove et al., 1997, 1998; Klein, 1999).  We argue that previous
conclusions on the conditions for cloud fragmentation have limited
applicability, since they did not allow for the nonlinear density and
velocity perturbations that are ubiquitous in molecular cloud cores.
Over the past year, we have begun to simulate the evolution of
marginally stable, turbulent cores. These models have radii, masses,
density contrasts, turbulent linewidths, and projected velocity
gradients consistent with observations of low-mass molecular cloud
cores.
Our models are evolved in time under
self-gravitational hydrodynamics with AMR using a barotropic equation
of state that models the transition from an isothermal to an adiabatic
equation of state.  We examine several properties of the resulting
protostellar fragments and discuss the qualitative nature of the
fragmentation process in realistic cloud cores:  the transition from
single to binary and multiple stars; the formation of misaligned binary
systems; and the role played by filament formation in the formation of
stars.  \end{abstract}

\section{Introduction}

Most stars exist in gravitationally bound binary and low-order multiple
systems (Bodenheimer et al 2000).  Moreover, the observed trend of
decreasing fraction of binary and multiple systems with age strongly
suggests that multiple star systems must have formed via fragmentation
during the earliest stages of cloud collapse, rather than from capture
of two or more stars formed individually (i.e. Simon \etal 1995).
Multiple star formation is central to our understanding of the star
formation process, and has enormous significance for the study of
formation of planetary systems as well.

While the results of previous authors (notably Hoyle 1953, Hunter 1967,
Low \& Lynden-Bell 1977, Silk 1977, Rees 1977, Larson 1985, and
Inutsuka \& Miyama 1992, among many others) have advanced the state of
our knowledge of the dynamics underlying the fragmentation process, no
theoretical fragmentation models to date have been able to convincingly
explain the observed properties of binaries.  Most analytic and
semi-analytic work has treated only the earliest phases of the
collapse, typically via a linearized analysis, and most numerical work
has treated only the earliest isothermal phase of the collapse, well
before most of the mass of the cloud core has been accreted onto the
fragments.  In addition one must also be able to follow the collapse
long enough to determine the number of protostars, their masses, and
their orbits.  Long before this point, the gas ceases to be
isothermal.  Hence, {\it a nonlinear, nonisothermal analysis is
absolutely essential}.

Until very recently, the extreme variations in length scale inherent in
collapse have made it difficult to perform accurate calculations of the
fragmentation process, which is intrinsically three-dimensional in
nature.  Our recent development of a robust adaptive mesh refinement
(AMR) self-gravitational hydrodynamics code has resulted in our
discovery of a physically motivated constraint on the resolution (the
Jeans condition) that must be satisfied to suppress artificial fragmentation
(Truelove \etal 1997).  This condition indicates that many published
solutions were underresolved and therefore gave misleading results.
Since our code is able to adhere to the Jeans criterion, it is ideally
suited for the treatment of the collapse and fragmentation of molecular
cloud cores.

Over the last year, we have begun a program of research investigating
the properties of marginally stable, turbulent low-mass molecular cloud
cores. These models have radii, masses, density contrasts, turbulent
linewidths, and projected velocity gradients consistent with
observations of low-mass molecular cloud cores.

\subsection {Key Questions to be Addressed}

Answers to basic questions concerning the fragmentation process remain
elusive:

{\bf {(1)}} Why do most stars form in binary systems?

{\bf (2)} What determines the efficiency of star formation, i.e., the
fraction of a core that will end up in protostellar objects?

{\bf (3)} What are the properties of protostellar disks formed in
turbulent molecular cloud cores? Are they typically aligned or
misaligned relative to one another?

{\bf (4)} What determines the distribution of fragment masses, which is
related to the initial distribution of stellar masses?  Similarly, what
determines the initial orbital angular momenta and rotational spins of
the individual fragments?

Our goal  is to make significant advances in our
understanding of all of these questions.  We shall study the collapse
and fragmentation of of turbulent molecular cloud cores with conditions
typical of nearby star-forming regions (Jijina \etal 1999), and
determine whether our models are consistent with the observed frequency
of binary star systems.  To the best of our knowledge, no previous
study of binary formation has considered either realistic
near-equilibrium clouds or realistic perturbation spectra.  This study
has the potential of leading to a major advance in our understanding of
how fragmentation proceeds in real molecular cloud cores.

\section {The Key Role of Nonlinear Perturbations : A Numerical
Experiment}

In order to reach more general conclusions on the outcome of the
collapse, several groups have attempted to establish a general
criterion as to whether an individual cloud model will fragment or
not.  Several classic studies were conducted based on the results of
early axisymmetric and 3D numerical models. Notably, Tohline, Durisen,
and McCollough (1985), Hachisu \& Eriguchi (1985) and Miyama (1992)
assumed that the collapse of rigidly rotating, uniform density spheres
resulted in a self-gravitating, rotating equilibrium isothermal disks
consisting of the entire mass of the cloud. By analyzing these disks
for linear stability to growing modes using the analysis of Goldreich
\& Lynden-Bell (1965), they arrived at the classic fragmentation
criterion for uniform cores that is still frequently cited in the
literature today.  Their criterion is stated in terms of the standard
two-dimensional parameter space of the cores: $\alpha = 5 c_s^2 R_0 / 2
G M $, which is proportional to the ratio of thermal to gravitational
energies within the core, and $\beta = \Omega_0^2 R_0^3 / 3 G M$, which
is proportional to the ratio of rotational to gravitational binding
energies within the core. In terms of $\alpha$ and $\beta$, the classic
criterion for fragmentation is $\alpha \beta < 0.12$, suggesting that
typical slowly rotating cores with $\alpha \sim 1$ and $\beta \sim
0.02$ should easily fragment.

More careful analytic work by Tsuribe \& Inutsuka (1999a,b; hereafter
TI) on the dynamics of a collapsing, rotating ellipsoid, including the
inward propagation of the rarefraction front from the boundary of the
cloud, revised the classic criterion for the fragmentation of uniform
density, rigidly rotating isothermal cores.  Their numerical
simulations included the effect of small white noise, bar-mode, and
ring-mode density fluctuations.  The TI criterion predicts that cores
in a state of marginal virial stability should {\it never} fragment.  They
report their criterion to be in in good agreement with their numerical
results.  The TI criterion clarifies almost three decades of
previous numerical and analytic work.

In summary, the classic $\alpha - \beta$ criterion, which predicted
ubiquitous fragmentation for slow rotators, is now known to be invalid.
Moreover, the TI criterion predicts ubiquitous single-star formation
for realistic cores near virial equilibrium. {\it Significantly, this
means that the problem of binary formation, even in idealized uniform
cores, has remained unsolved to date.}

However, TI's results are limited in that their choice of
perturbation spectrum is not physically motivated; in particular, none
of the chosen perturbation spectra are consistent with the
linewidth-size relation observed in molecular cloud cores (Larson,
1981).  More important, the amplitudes chosen for the density
perturbations are unrealistically small: typical cores are observed to
have turbulent linewidths comparable to the isothermal soundspeed
(Jijina \etal 1999), and so the perturbations present in real molecular
cloud cores will typically be nonlinear.

The outcome of the initial collapse phase hinges upon whether the
timescale for the overall collapse of the core is greater or less than
the timescale for the growth of perturbations. When the timescale to
grow the perturbation is greater than that of the overall collapse, a
single fragment is formed; in the opposite case, a binary or multiple
system may be formed. Nonlinear perturbations play a key role in this
process, since the timescale for the gravitational instability to grow
a nonlinear perturbation is {\it always} less than that for a linear
perturbation.

To test the sensitivity of the TI fragmentation criterion to the choice
of perturbation amplitude, we apply an $m = 2$ density perturbation to
a variety of uniform clouds. We find that a modest increase increase in
the amplitude of the applied spectrum can cause models significantly
removed from their fragmentation transition curve to fragment. In
figure 1, we show such a model system with $\alpha = 1.0$, $\beta =
0.1$, and an applied amplitude of $25 \%$, which produced a binary.  In
contrast, the same background model run with a slightly smaller
perturbation amplitude of $10 \%$ produced a single object at a
comparable compression.

\begin {figure}
%  \plotone {klein.fig1.eps}
  \caption {An isothermal uniform model with
	    $\alpha = 1.0$, $\beta = 0.1$, with 25\% $m = 2$ density
	   perturbation, shown after a compression
	    of over seven orders of magnitude in density. Note the
	    formation of a binary system.}
\end {figure}

{\it This result illustrates that nonlinear perturbations, which are
typically found in molecular cloud cores, play a critical role in
fragmenting uniform isothermal spheres close to marginal virial
stability.} We turn our attention in the next section to a more
realistic choice of background core models and applied spectra.

\section {Turbulent Low-Mass Molecular Cloud Cores: Initial Conditions}

The statistics of prestellar versus embedded stellar molecular cloud
cores suggest that the prestellar phase is greater than the free-fall
time of the core; hence, the core should be in an approximate state of
equilibrium (Beichman \etal 1986, Myers \etal 1995, Lee \& Myers,
1999). Accordingly, the initial conditions for a typical model we
simulate consist of self-consistent hydrodynamic equilibria of
turbulent, marginally stable, isothermal molecular cloud cores.  In
contrast, nearly all calculations that appear in the literature deal
with clouds unrealistically initially far from equilibrium (e.g. Boss
\& Bodenheimer, 1979, Boss 1991, Burkert \& Bodenheimer, 1993), leading
to highly supersonic collapse velocities.

We perturb the cloud using Gaussian spectral perturbations on the
velocity field with a $k^{-4} d^3k$ spectrum consistent with Larson's
linewidth-size relation (Larson, 1981; Dubinski, Nararyan, \& Phillips
1995).  A limitation of our present procedure is that the density
perturbations form only after the velocity perturbations have time to
evolve; by the time the density perturbations have formed, the
turbulence has decayed somewhat.

The Bonnor-Ebert values for mass and radius are rescaled to account for
turbulent support :  \begin {equation} M \rightarrow M (1 + {\cal M}^2
)^{3/2} , \end {equation} \begin {equation} R \rightarrow R (1 + {\cal
M}^2 )^{1/2} , \end {equation} where ${\cal M} = \sigma / c_s$ and
$\sigma$ is the 1D nonthermal velocity dispersion. We take ${\cal M}$
as the first dimensionless parameter specifying our model cores.

As the cloud collapses, the heating rate due to compression rises,
while the ability of the gas to cool is impeded by greater opacity due
to the higher column density of dust.  As a result, the collapse
evolves from isothermal to quasi-adiabatic.  To follow this evolution
in this current set of calculations, we use a barotropic equation of
state that transitions from isothermal to polytropic at a
characteristic density of $\rho_{\rm crit}$, i.e., \begin {equation} P
(\rho) = c_{\rm iso}^2 \rho + K_{\rm stiff} \rho^{\gamma_{\rm stiff} },
\end {equation} where $P$ is the gas pressure as a function of density
$\rho$, $c_{\rm iso}$ is the isothermal soundspeed, $\gamma_{\rm
stiff}$ is the adiabatic exponent in the optically thick regime, and
$K_{\rm stiff}$ is a parameter fixed by the requirement that the
isothermal and the adiabatic pieces of the equation of state are equal
at a characteristic density $\rho_{\rm crit}$:  \begin {equation}
K_{\rm stiff} = { \rho_{\rm crit}^{1 - \gamma_{\rm stiff} } \over
c_{\rm iso}^2} \end {equation} (Boss \etal 2000; Klein \etal 1999).  We
take the density ratio $\rho_{\rm crit} / \rho_{\rm edge}$ as our
second dimensionless parameter.  As Boss \etal\ (2000) point out, this
barotropic equation of state is very approximate, but it enables us to
take a first step in understanding what role non-isothermal effects
play in the more general case.

Hence, our barotropic models are completely specified in a
two-dimensional parameter space: (${\cal M}$, $\rho_{\rm crit} /
\rho_{\rm edge}$). Both of these parameters are constrained by
observation. Nearby star-forming low-mass cores have typical values of
${\cal M} \sim 0.5 - 1.5$ (Jijina \etal 1999).  We choose a fiducial
value of the external density of $\rho_{\rm edge} = 5 \times 10^{-20}$
gm cm$^{-3}$ and a critical density ratio in the range $\rho_{\rm
crit}/\rho_{\rm edge} \sim 10^4 - 10^6$. Following Masunaga \& Inutsuka (1998),
the critical density ratio
can be related to the Planck mean opacity $\kappa_P$ by noting that
an optically thick, gravitationally bound
structure at $\rho_{\rm crit}$ will have a 
characteristic scale of about the local Jeans length $\lambda_J$. 
Hence,
\begin {equation}
  \kappa_P \rho_{\rm crit} \lambda_J \sim 1
\end {equation}
With some rearrangement, and assuming we have a cosmic gas mix, we find
\begin {equation}
  \kappa_P \sim 2 \cdot 10^{-1} {\rm cm^2/g} ({5 \cdot 10^{-20} {\rm gm/cm^3} \over \rho_{edge} } )^{1/2} ({10^5 \over \rho_{\rm crit} / \rho_{\rm edge} })^{1/2}
({T_{init} \over 10 {\rm K} })^{-1/2}
\end {equation}
Values of $\kappa_P$ appropriate for cold, dense molecular cloud cores
are model-dependent on dust grain composition, 
carbon and silicate inclusions, icy mantles,
and dust size distribution in the presence of sticking and destructive effects.
Ranges $\kappa_P \sim$ 0.02 - 0.2 cm$^2$/g are cited in the literature
(Preibisch \etal 1993).
This implies a range of $\rho_{\rm crit} / \rho_{\rm edge} \sim
10^5 - 10^7$.

Below we explore the
effect of turbulence by fixing $\rho_{\rm crit} / \rho_{\rm edge} =
10^4$, and vary ${\cal M}$ from $\sim 0.5 - 3.0$. This choice of
the stiffening parameter
is slightly too stiff in comparison to realistic dust grain models. 
However, the 
enhanced stiffening allows us to evolve our calculations much longer in
accretion clock time. A future set of calculations will relax this assumption,
and will evolve with a relatively soft equation of state for comparison.

The principal physical processes that we are neglecting in this study
are those associated with magnetic fields.  Although ambipolar
diffusion diminishes the importance of the magnetic field relative to
both gas pressure and gravity as the collapse proceeds (Ciolek \&
Mouschovias, 1994), the role of the field in the dynamical processes
underlying fragmentation is almost completely unknown.  However, in
order to fully understand the dynamics of the collapse, it is
worthwhile to carry out careful unmagnetized as well as magnetized
studies; our current program of research represents a step in this
direction.  We also note that some recent studies have suggested that
the magnetic field may be less important than previously believed
(Padoan \& Nordlund 1998).

\section {Results}

We have conducted a preliminary survey of the parameter space of
turbulent cores by holding the equation of state fixed (i.e., setting
$\rho_{\rm crit} / \rho_{\rm edge}=10^4$) while varying the turbulent
Mach number ${\cal M}$ over the range of typical values seen in
low-mass cores (median ${\cal M} \sim 0.5 - 1.5$, with the low and high
values corresponding to Taurus and Orion, respectively), to small
cluster-forming clumps (${\cal M} \sim$ several) (Jijina \etal 1999).

All calculations reported utilize our AMR code, which obeys the
local Jeans condition by resolving the local Jeans length with at least
four cells at all points in space and time (Truelove \etal 1997;
Truelove \etal 1998). Our initial, unperturbed, centrally condensed
equilibrium isothermal spheres are chosen to be very slightly
subcritical, with a density contrast of edge to center of 10, and a
mass of $99.9\%$ that of the critical Bonnor-Ebert value. As a result,
without turbulence, the models remain stationary indefinitely.  In
order to obtain physical values from our dimensionless models, we set
the density at the edge of the cloud at $\rho_{\rm edge}=5\times
10^{-20}$ g cm$^{-3}$ as remarked above and the isothermal soundspeed
at $c_{\rm iso} = 1.8 \times 10^4$ cm s$^{-1}$ (corresponding to $T
\simeq 10$ K and cosmic abundances).  A fiducial turbulent core with
${\cal M} = 1.0$ then has a radius of $R = 0.07$ pc and a mass of $M =
2.9 M_{\odot}$.  The characteristic timescale for the evolution of the
entire cloud is the free-fall time evaluated at the edge density,
$t_{\rm ff}\equiv(3\pi/32 G\rho_{\rm edge})^{1/2} =3.0\times 10^5$ yr.
These fiducial choices are typical of nearby star-forming regions,
though a substantial variance exists within in each molecular cloud,
and from cloud to cloud.

\subsection {${\cal M} = 0.58$}

This model corresponds roughly to the median level of turbulence seen
in the Taurus molecular cloud, which has the lowest level of turbulence
of any of the nearby star-forming molecular clouds (Jijina \etal
1999).  In the fiducial choice of edge density and sound speed
specified above, this model has $R = 0.06$ pc and $M = 1.58
M_{\odot}$.  A single fragment is formed, surrounded by a centrifugally
supported protostellar disk (see Fig. 2).  By the end of the
calculation, which was run for a time $\sim 0.5 t_{\rm ff}$ after the
time of formation of the fragment, the total amount of mass in the
fragment is roughly half of the initial core mass. This is consistent
with a nearly constant mass accretion rate over this time interval.

\subsection {${\cal M} = 1.7$}

In the opposite extreme of turbulence seen in nearby molecular clouds
is Orion, which has a median value ${\cal M} \sim 1.5$ (Jijina \etal
1999). We ran a calculation with turbulent level of support consistent
with this value, ${\cal M} = 1.7$. Using the fiducial edge density and
sound speed, we find this model has $R = 0.1$ pc and $M = 8.2
M_{\odot}$.  It appears that a binary system has formed, which by the
endpoint of the calculation has accreted about $10\%$ of the total mass
of the core. The binary system is seen to emerge along the ends of a
filamentary structure in the turbulent flow (see Fig. 3). Simple
estimates of the mass flux through the filamentary structure reveal
that the density within the filament is typically one-several orders of
magnitude greater than the ambient medium density, and hence that
virtually all mass flux occurs along filaments.  Both of the members of
the binary appears to have formed a protostellar disk.  Moreover, the
disks appear to be misaligned relative to one another, as Smith \etal
(Smith \etal 1999) suggest is the case for T Tau.  We note that at
this point in the evolution, the binary has a very wide separation,
$\sim 10^4$ AU.  It is possible that this system will evolve into a
much closer binary, but at present this is too computationally
demanding to determine.

\begin {figure}
%  \plotone {klein.fig2.eps}
  \caption {The log of the projected column density along the
	    $z$-axis of an ${\cal M} = 0.58$, $\rho_{\rm crit} /
	    \rho_{\rm edge} = 10^4$ run. A single fragment has formed,
	    and followed to the point it has accreted about 50\% of the
	    total mass of the core.}

%  \plotone {klein.fig3.eps}
  \caption {The log of the projected column density along the
	    $z$-axis of an ${\cal M} = 1.73$, $\rho_{\rm crit} /
	    \rho_{\rm edge} = 10^4$ run. A binary system is seen to
	    emerge here, which has accreted about 10\% of the total
	    mass of the core.}
\end {figure}

\subsection {${\cal M} = 2.9$}

Lastly, extending our models into the range of more turbulent clumps
that can form clusters of stars, we evolved one model with ${\cal M} =
2.9$. Using the fiducial edge density and sound speed, this model has
$R = 0.16$ pc and $M = 29 M_{\odot}$.  A complex network of shocked
filamentary structures rapidly evolves, and at least six fragments
emerge from the flow. However, due to computational constraints, we
have been able to evolve this model only to the point that $0.1 \%$ of
the total mass of the core has been accreted.  The fragments appear to
form {\it within} the filamentary structures, at those locations in the flow
where filaments cross or bend.  We are examining the possibility that
the mass accretion flow in the filaments is largely responsible for the
growth of the protostellar fragments for such higher Mach cases, though
the duration of this calculation is too short to draw firm
conclusions.

\section {Conclusions}

Beginning with models that have density contrasts, linewidths, radii,
masses, and projected linewidth gradients consistent with observations
of low-mass molecular cloud cores, we have demonstrated that turbulence
plays a critical role in their fragmentation. Cores that are
significantly subsonic do not fragment in our models, whereas transonic
and supersonic cores typically do.  Since we inject
turbulence on a level consistent with what is actually observed, our results
suggest that turbulent fragmentation is likely to play a critical role
in the formation of binary and multiple systems in molecular cloud
cores.  Our preliminary calculations suggest that filament formation
is ubiquitous in turbulent cores.  The protostellar cores
appear to form within the filamentary structure, and the filaments
themselves  may be the principal
conduit for mass transfer onto the protostellar disk. 

However, several important issues remain to be resolved. First, our
models have difficulty producing binary systems for low turbulent Mach
numbers (i.e. ${\cal M} < 0.5$), whereas observations in Taurus
indicate no diminution in the number of binaries in this range of $\cal
M$.
Second, all models must be evolved for a global freefall timescale
in order to conclusively answer questions concerning the outcome of
fragmentation in the cores. Lastly, many different turbulent
realizations must be used in order to gather statistics regarding final
orbital separation and mass ratio, which must also be checked against
observation. We plan to address all of these issues in a subsequent
paper (Fisher \etal 2000). This gauntlet of tests provides us an unprecedented
opportunity to examine the fragmentation paradigm in the context of
realistic models.

\acknowledgments{This work is supported in part by a grant from the NASA ATP
program to
the Center for Star Formation Studies.  The research of RIK is supported
in part  under
the auspices of the US Department of Energy at the
Lawrence Livermore National Laboratory
under contract W-7405-Eng-48.  The research of CFM is
supported in part by NSF grant AST95-30480. RTF is supported in part by
the NASA GSRP program. We thankfully acknowledge the use of the NSF
NPACI high performance computing facility; all calculations reported
here were performed on the NPACI T90.}

\begin {references} \reference Beichman, C., Myers, P., Emerson, J.,
Harris, S., Matthieu, R., Benson, P., \& Jennings, R. 1986, ApJ, 307,
337.

\reference Bodenheimer, P., Burkert, A., Klein, R. I., \& Boss, A. P.
2000, in Protostars and Planets IV, ed. V. Mannings, A. P. Boss, \& S.
Russell (Tucson: University of Arizona Press), p.  675.

\reference  Boss, A.~P., Fisher, R.~T., Klein, R.~I., McKee, C.~F.,
2000,ApJ, 528, 325.

\reference Boss, A., \& Bodenheimer, P., 1979, ApJ, 234, 289.

\reference Boss, A.~P. 1991, Nature, 351, 298.

\reference Burkert, A. \&  Bodenheimer, P. 1993, MNRAS., 264, 798.

\reference Ciolek, G. \& Mouschovias, T. 1994, ApJ, 425, 142.

\reference Dubinski, J, Narayan, R., \&  Phillips, T. 1995, ApJ, 448,
226.

\reference Fisher, R., Klein, R., \& McKee, C. 2000, to appear in ApJ.

\reference Goldreich, P., \& Lynden-Bell, D. 1965, MNRAS, 130, 97.

\reference Hachisu, I., \& Eriguchi., Y. 1985, A \& A, 143, 355.

\reference Hoyle, F. 1953, ApJ, 118, 513.

\reference Hunter, C. 1967, ApJ, 136, 594.

\reference Inutsuka, S. \& Miyama, S. 1992, ApJ, 388, 392.

\reference Jijina, J., Myers, P., \& Adams, F. 1999, ApJS, 125, 161.

\reference Klein, R.~I. 1999, JCAM, 109, 123.

\reference Klein, R.~I., Fisher, R.~T., McKee, C.~F., \&  Truelove,
J.~K. 1999,  in Numerical Astrophysics 1998, eds.  Miyama, S., \&
Tomisaka, K.

\reference Larson, R. 1981, MNRAS, 194, 809.

\reference Larson, R. 1985, MNRAS, 214, 379.

\reference Lee, C., \&  Myers, P., ApJS, 123, 233.

\reference Low, C., \& Lynden-Bell, D., MNRAS, 176, 367.

\reference Masunaga, H. \& Inutsuka, S. 1998, ApJ, 495, 346.

\reference Miyama, S. 1992, PASJ, 44, 193.

\reference Myers, P., Bachiller, R., Caselli, P., Fuller, G, Mardones,
D., Tafalla, M., \& Wilner, D. 1995, ApJ, 319, 340.

\reference Padoan, P., \& Nordlund, A. P. 1998, ApJ, 504, 300.

\reference Preibisch, T., Ossenkopf, V., Yorke, Y., \& Henning, T. 1993,
A \& A, 279, 577.

\reference Rees, M. 1977, MNRAS, 176, 483.

\reference Silk, J. 1977, ApJ,214,152.

\reference Simon, M., Ghez, A., Leinert, C., Cassar, L., Chen, W.,
Howell, R., Jameson, R., Matthews, K., Neugebauer, G., \& Richichi, A.
1995, ApJ, 443, 625.

\reference Smith, D., Wood, K., Whitney, B., Kenyon, S., and Stassun,
K.  1999. AAS, 195, 210.

\reference Tohline, J., Durisen, R., \& McCollough, M. 1985. ApJ,
 298, 220.

\reference Truelove, J.K., Klein, R.I., McKee, C.F., Holliman, J.H.,
Howell, L. H., Greenough, J.A., \& Woods, D. T., 1997. ApJ, 489, L179.

\reference Truelove, J.K., Klein, R.I., McKee, C.F., Holliman, J.H.,
Howell, L.H., Greenough, J.A., 1998. ApJ, 495, 821.

\reference Tsuribe, T., \& Inutsuka, S. 1999. ApJ, 526, 307.

\reference Tsuribe, T., \& Inutsuka, S. 1999. ApJ, 523, 155.

\end {references}

\end{document}